\documentclass[aps,prd,twocolumn]{revtex4}
\usepackage{latexsym}
\usepackage{epsfig}
\newcommand{\del}{\partial}
\newcommand{\beq}{\begin{eqnarray}}
\newcommand{\eeq}{\end{eqnarray}}
\newcommand{\be}{\begin{eqnarray}}
\newcommand{\ee}{\end{eqnarray}}
\newcommand{\bk}{{\bf k}}

\newcommand{\bx}{{\bf x}}

\newcommand{\ra}{\rightarrow}

\newcommand{\nn}{\nonumber}

\begin{document}

\preprint{astro-ph/0409163}
\title{The CMB Spectrum in Cardassian Models}
\date{\today}

\author{Tomi Koivisto}
\email{tomikoiv@pcu.helsinki.fi}
\affiliation{Helsinki Institute of Physics,FIN-00014 Helsinki, Finland}
\affiliation{Department of Physics, University of Oslo, N-0316 Oslo, Norway}
\author{Hannu Kurki-Suonio}
\email{hkurkisu@pcu.helsinki.fi}
\affiliation{Department of Physical Sciences, University of Helsinki, FIN-00014 Helsinki, Finland}
\author{Finn Ravndal}
\email{finn.ravndal@fys.uio.no}
\affiliation{Department of Physics, University of Oslo, N-0316 Oslo, Norway}

\begin{abstract}

The dark energy in the Universe is described in the context of modified
Friedmann equations as a fluid parameterized by the density of dark matter and
undergoing an adiabatic expansion. This formulation is applied to the
Cardassian model. Choosing then parameters consistent with the supernova
observations, it gives a background expansion in which the cosmic temperature
fluctuations are calculated. The resulting spectrum is quite similar to what is
obtained in the standard concordance model. If the Cardassian fluid is
interpreted as a new kind of interacting dark matter, its overdensities are
driven into oscillations when the interaction energy is rising in importance.
This does not occur in a description of the Cardassian fluctuations motivated
by theories of modified gravity. There the energy of the underlying matter is
also conserved, which requires appearance of effective shear stress in the
late Universe. In both approaches that allow fluctuations the thermal power
spectrum at large scales is much too strongly enhanced by the late
integrated Sachs-Wolfe effect. With the interacting dark matter 
assumption, we conclude that the Cardassian model is ruled out by observations, 
expect in a small neighbourhood of the $\Lambda$CDM limit.

\end{abstract}

\maketitle

\section*{Introduction}

Observations of distant supernovae indicate that the expansion of the Universe now undergoes an acceleration\cite{Riess:2004nr,Riess:1998cb,
Perlmutter:1998np}.
This is also consistent with the most recent and accurate measurements of the fluctuations in the cosmic microwave background by the WMAP
satellite\cite{Spergel:2003cb}. These also imply that the Universe is at least very nearly spatially flat.

A possible explanation of these observations requires that the Universe
contains a new and unknown component, called dark energy, in addition to
radiation, baryons and dark matter. This could be Einstein's cosmological
constant or a more dynamical component based on the universal presence of a
scalar field called quintessence\cite{Zlatev:1998tr,Caldwell:1997ii}. Alternatively, one could
contemplate the possibility that standard Einstein gravity is not valid at very
large scales so as to allow for modifications of the Friedmann
equations\cite{Dvali:2000hr,Carroll:2003wy}.

We will here consider one particular such proposal by Freese and
Lewis\cite{Freese:2002sq} where the total energy density is made up of just the
usual radiation, baryons and cold dark matter densities $\rho_r$, $\rho_b$ and
$\rho_c$  but they enter in a non-linear way in the Friedmann equation for the
scale parameter $a(t)$ in a spatially flat Universe, \be
       H^2  = {8\pi G\over 3}g(\rho_r,\rho_b,\rho_c)
\ee
where $H = \dot{a}/a$ is the Hubble parameter. There are several such proposals based on ideas from brane physics\cite{Chung:1999zs} or some unknown
interactions between matter particles\cite{Gondolo:2002fh}. But in order to make contact with the CMB physics, we will define the model so that
the Friedmann equation takes the form
\beq
       \left({\dot{a}\over a}\right)^2 = {1\over 3M^2}(\rho_r + \rho_b +\rho_K)                    \label{Fried}
\eeq
where $M= (8\pi G)^{-\frac{1}{2}}$ is the reduced Planck mass and
 \beq
        \rho_K = \rho_c[1 + B\rho_c^{-q\nu}]^{1/q}                            \label{cardass}
 \eeq
is the modified polytropic Cardassian energy density. Here $B$ is a positive
constant, $q$ is the polytropic index and it is modified as long as the
parameter $\nu \neq 1$. All the unknown physics is now lumped into the dark
sector. At early times when the matter density $\rho_c$ is large, the last term
in (\ref{cardass}) will be negligible and $\rho_K \approx \rho_c$. However, at
late times when $\rho_c$ becomes small, the last term dominates and will act as
a dark energy component due to unknown properties of dark matter. The energy
fractions $\Omega_i = \rho_i/\rho_{cr}$ with $\rho_{cr} = 3M^2H^2$, are seen
from (\ref{Fried}) to satisfy $\Omega_r +\Omega_b +\Omega_K = 1$ since there is
no curvature in space. In our calculations we use the WMAP values\cite{Spergel:2003cb}
for these parameters. Then $\Omega_c = 0.226$, $\Omega_b = 0.044$, and
$\Omega_r$ we get from assuming three massless neutrino species and the
background temperature $T=2.726~\mbox{K}$. For $h$ we use the value $0.72$.
Today we will therefore have $\Omega_{K0} =1$ when we neglect the small
contributions from radiation and baryons. Given $q$ and $\nu$, the remaining
parameter $B$ follows then from
 \be
        B = \rho_{cr}^{\,q\nu}\big(\Omega_{c0}^{-q} - 1\big)\Omega_{c0}^{\,q\nu}.
 \ee
Ordinary dark matter has zero pressure and satisfies the conservation equation
$\dot{\rho}_c + 3H\rho_c = 0$ (in the homogeneous and isotropic background
Universe). Thus we have $\rho_c =\rho_{c0}/a^3$, and correspondingly for
the other standard ingredients, with $a=1$ today. The Friedmann equation
(\ref{Fried}) can then be integrated to give the full background
evolution of the scale parameter $a(t)$ with Cardassian energy present.

\begin{figure}[t]
\begin{center}
\includegraphics[width=0.45\textwidth]{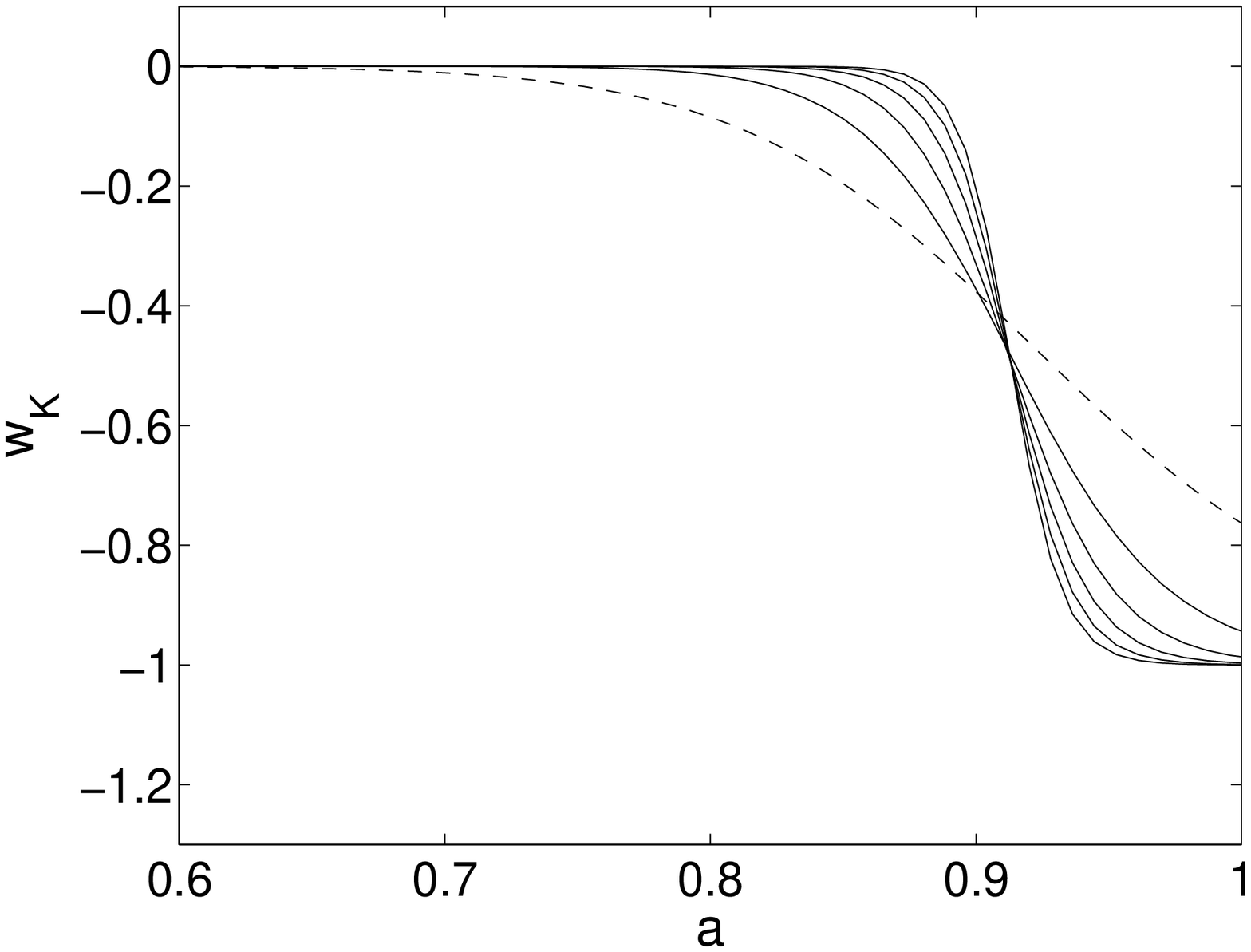}  \label{equation of state}
\includegraphics[width=0.45\textwidth]{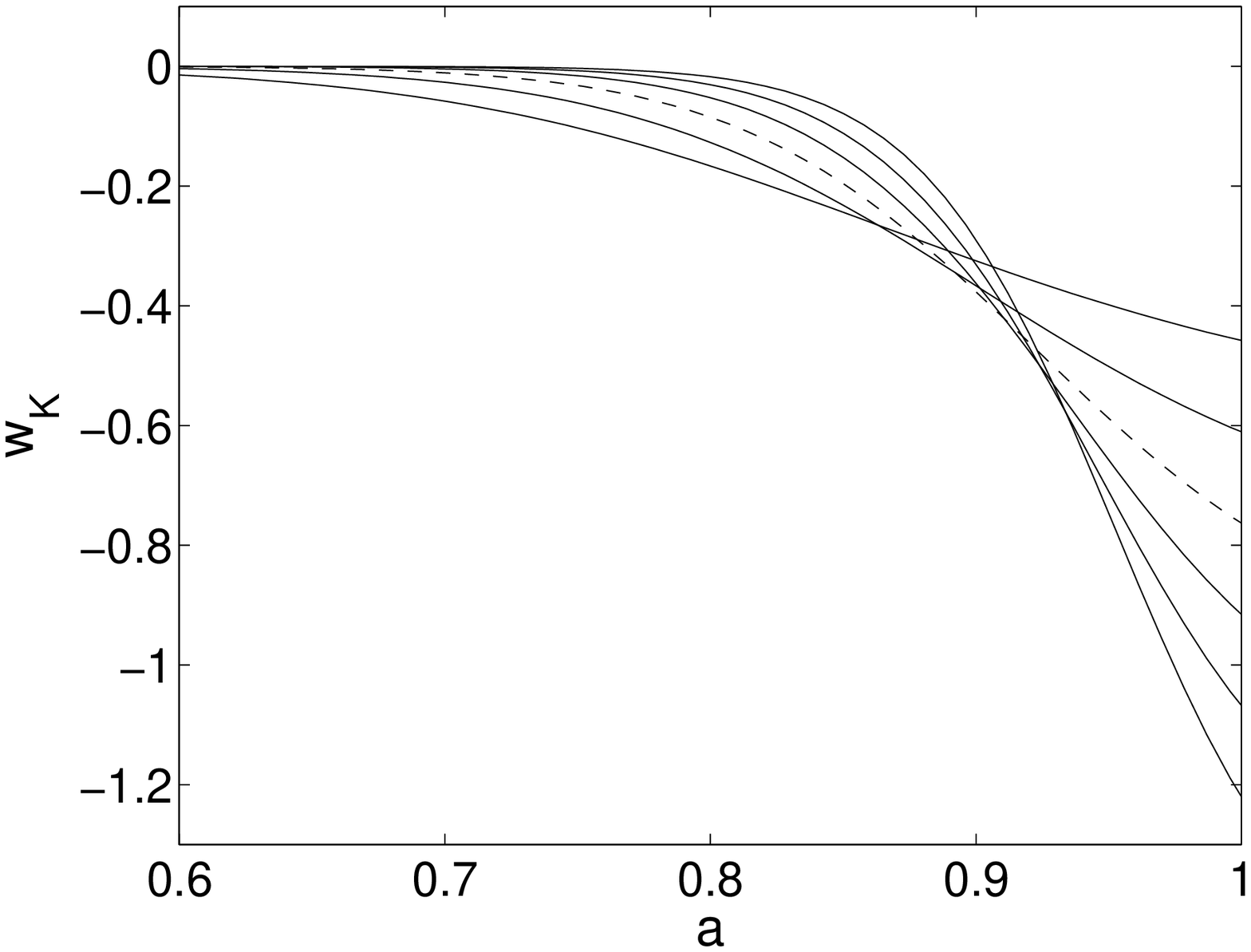}
\caption{\label{eosfig}Equation-of-state parameter $w_K$ as a function of the scale factor. The upper panel has $\nu =1$ and
$q$-values $q=1$,$2$,$3$,$4$,$5$ and $q=6$ (the steepest curve is for $q=6$). The lower panel has $q=1$ and
$\nu=0.6$,$0.8$,$1.0$,$1.2$, $1.4$ and $\nu=1.6$ (the most negative $w_K$ today is for $\nu=1.6$). The dashed curves
correspond to $\Lambda$CDM.}
\end{center}
\end{figure}

Following Gondolo and Freese\cite{Gondolo:2002fh}, we will assume that
the Cardassian energy density $\rho_K$ is that of a fluid which undergoes an
adiabatic expansion when the Universe evolves. It will thus have a pressure
$p_K$ which can be obtained from its conservation equation $\dot{\rho}_K +
3H(\rho_K + p_K) = 0$. Eliminating the Hubble parameter $H$ between this and
the conservation equation for the dark matter, the Cardassian pressure will
follow from \beq         \label{pressure}
            p_K = \rho_c {d\rho_K\over d\rho_c} - \rho_K.
\eeq
For the choice (\ref{cardass}) it gives
\beq
            p_K = - \nu B\rho_c^{-q\nu+1}\Big[1 +  B\rho_c^{-q\nu}\Big]^{{1\over q} - 1}.              \label{press}
\eeq
This equation of state can be written on the standard form $p_K = w_K\rho_K$ with
\beq
            w_K = {-\nu B \rho_c^{-q\nu}\over 1 + B\rho_c^{-q\nu}}.                \label{w_K}
\eeq
At late times when $\rho_c \ra 0$ we see that $ w_K \ra -\nu$. We therefore expect late-time
acceleration when the parameter $\nu > 1/3$. The equation-of-state parameter $w_K$ is zero at early times,
and approaches the value $w_K=-\nu$ in the late universe, more rapidly for larger $q$. Late evolution
of $w_K$ for different values of the parameters $q$ and $\nu$ is shown in Fig.\ref{eosfig}.

The late expansion history of these models has been compared to the observational
data of supernovae\cite{Zhu:2003sq,Wang:2003cs} and of other astrophysical objects\cite{Zhu:2002yg,Zhu:2003wv}. The
locations of acoustic peaks in the CMB spectrum has also been considered before\cite{Sen:2003cy,Frith:2003ad}.
We will calculate the full CMB spectrum in these models, first assuming that the
Cardassian component accounts only for the background expansion of the
Universe. This would, however, correspond to a time-variable but nonfluctuating
vacuum energy, to which we have no physical motivation\footnote{See however the mention
of an interacting vacuum energy in the discussion of case II.}. This is our case I.

A more realistic description, which allows fluctuations in the Cardassian fluid, is our case II. Then
the Cardassian fluid can be thought of as an interacting dark matter. The
energy density\footnote {Note that this would correspond to $\rho_K$ in the
notation of\cite{Gondolo:2002fh}.} $\rho_K-\rho_c$ would then be attributed to an
unknown particle species or field mediating the dark matter interactions. Thus
$\rho_c$ does not satisfy an independent energy conservation law.  It only
satisfies a mass conservation law, whereas we have energy conservation for the
total $\rho_K$.

In our case III we impose energy conservation separately for the cold dark
matter. This can be motivated by theories of modified gravity. In these theories,
we can generally write the Einstein equations as
equations as \be
   G_{\mu\nu} = \frac{1}{M^2}(T_{\mu\nu}^M + T_{\mu\nu}^C),
\ee
where $M$ is for matter and $C$ for the corrections to the standard gravity. The latter can be
parameterized as a function of the matter energy density.
We illustrate this with the example of corrections in the form of DGP gravity\cite{Dvali:2000hr},
\beq
    H^2 \pm \frac{H}{r_0} = \frac{1}{3M^2}\rho_c. \label{dgpgrav}
\eeq
This can rewritten as
\beq \label{dgppres}
   3M^2H^2 = \rho_c + \frac{3M^2}{2r_0^2} \mp \left(\frac{9M^4}{4r_0^4} + \frac{3M^2}{r_0^2}\rho_c\right)^\frac{1}{2}. 
\eeq
The RHS we then interpret as an energy density of a fluid. Granted that the matter conservation also now applies as
usually, we can find the pressure of this fluid
 \be
    p & = & -2M^2[\frac{3}{2}\dot{H} + H^2] \\
    & = & \mp \frac{\rho_c}{\left(1+ \frac{4r_0^2}{3M^2}\rho_c\right)^\frac{1}{2}} -
     \frac{3M^2}{2r_0^2}
     \pm \left(\frac{9M^4}{4r_0^4} + \frac{3M^2}{r_0^2}\rho_c\right)^\frac{1}{2}, nn
 \ee
which gives the equation of state for this effective matter. It would also follow
by using Eq.(\ref{pressure}) with Eq.(\ref{dgppres}). However, in this
paper we consider only Cardassian modifications to the Friedmann equation. For
a recent investigation of the cosmological constraints for modifications
generalized from Eq.(\ref{dgpgrav}), see Elgar\o y and
Multam\"aki\cite{Elgaroy:2004ne}. Note also that we have included only the
contribution from the density of dark matter in Eq.(\ref{dgpgrav}),
because the corrections to the Einstein gravity do not become
important until the matter dominated era. This justifies our use of
Eq.(\ref{Fried}) also in the case III\footnote{We expect that including baryon
contribution in $\rho_K$ would not change the results significantly.}.

\section*{Case I: No Cardassian fluctuations.}

\begin{figure}[th]
\begin{center}
\includegraphics[width=0.45\textwidth]{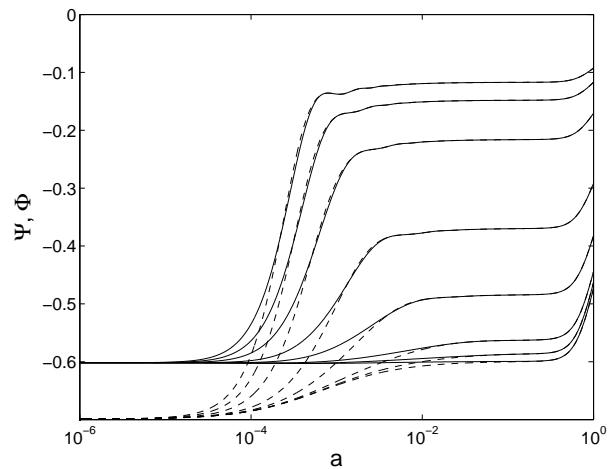}
\caption{\label{psi1}Evolution of $\Psi$ (solid line) and $\Phi$ (dashed line) in a Cardassian background (case I).
From bottom to top, $k = 0.00024, 0.0012, 0.0024, 0.0060, 0.012, 0.036,0.048$ Mpc$^{-1}$.}
\end{center}
\end{figure}

In order to calculate the temperature fluctuations in this Universe, we now consider these cases. Firstly we assume that the
Cardassian fluid just provides a modified background for the evolution without any internal fluctuations. These take place in the ordinary components
of radiation, baryons and dark matter and can be calculated by standard methods for any values of $q,\nu > 0$. In Fig.\ref{psi1} we show the Bardeen
potentials
for the choice $\nu = 0.8$ and $q = 1.5$ and for different values of the wave number $k$ which sets the scale of the
fluctuations.\footnote{Our $\Psi$ ($\Phi$) is the $\Psi$ ($-\Phi$) of Kodama and Sasaki\cite{Kodama:1985bj}, which is $\psi$ ($\phi$) of Ma and
Bertchinger\cite{Ma:1995ey} and $\Phi_A$ ($-\Phi_H$) of Bardeen\cite{Bardeen:1980kt}.} The evolution of gravitational potentials is seen to be very
similar
to what is found in $\Lambda$CDM models on all scales.

\begin{figure}[th]
\begin{center}
\includegraphics[width=0.45\textwidth]{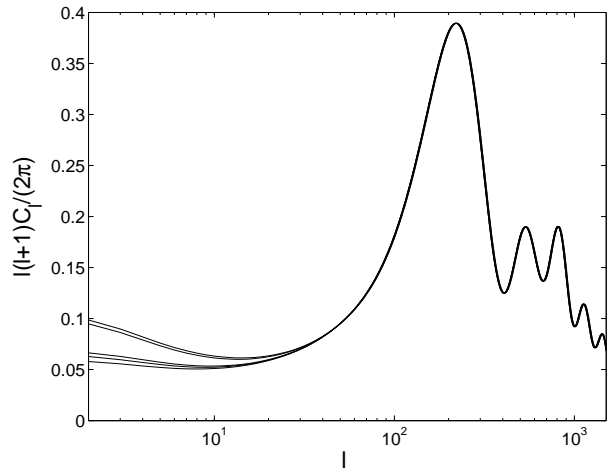}
\caption{\label{spectra1}CMB spectra in a Cardassian background (case I). From bottom to up, $(q,\nu)=(1,1),(1.5,0.8),(2,0.7),(10,0.6)$ and
$(100,0.6)$. The bottom curve with both $q$ and $\nu$ equal to 1 corresponds to the $\Lambda$CDM model.}
\end{center}
\end{figure}

We have calculated the CMB spectra using the fluid approximation for photons
and the well known analytical result for the diffusion damping scale (e.g.
\cite{Dodelson}). We have tested the fluid approximation with various models
and found it to be in agreement with more exact calculations (using e.g.
CMBFAST\cite{Seljak:1996is}) within 5 per cent for $\ell$ smaller than a few hundred.
 This is sufficiently accurate to uncover the interesting large scale features
of the models we are considering, as is the main purpose of this investigation.
For simplicity, we do not include reionization. Our perturbations are normalized such that
the primordial curvature perturbation $\mathcal{R}=1$.

In Fig.\ref{spectra1} we show the full thermal spectrum obtained from this modified background. Here we use the same Cardassian parameters $(q,\nu) =
(1,1), (1.5,0.8), (2,0.7), (10,0.6)$ and $(100,0.6)$ which have been found by Savage, Sugiyama and Freese\cite{Savage:2004zv} to be consistent with
the SNIa observations and the age of the Universe. The peaks in the thermal spectra are located at the right places and the contribution
from the integrated Sachs-Wolfe effect is seen to increase modestly with $q$. However, by lowering $q$ below $1$ one cannot reduce the ISW effect
without losing the agreement of the spectrum with WMAP observations on smaller scales.

\section*{Case II: CDM fluctuations driven by the Cardassian fluid.}

As our second approach, we include fluctuations in the Cardassian fluid.
These will then drive the fluctuations in the underlying dark matter. In our
discussion of perturbations we adopt the notation of Ma and
Bertschinger\cite{Ma:1995ey}, with the only exception of using uppercase letters for
the gravitational potentials in the conformal-Newtonian gauge. In this gauge the
line element is then written as
 \be
      ds^2 =  a^2(\tau)[-(1+2\Psi)d\tau^2 + (1-2\Phi)\delta_{ij}dx^idx^j]. \quad
 \ee
We assume adiabatic perturbations, which means that the relations (\ref{cardass})
and (\ref{press}) between $\rho_K$, $p_K$, and $\rho_c$ will hold also for the
perturbed quantities.

Assuming no anisotropic stress in the Cardassian fluid, we find that its
density perturbations in this gauge obey the equation of motion
 \beq
       \delta'_K = (1 + w_K)(-\theta_K + 3\Phi') +
       3\mathcal{H}(w_K -  c_K^2)\delta_K  \label{fluct1}
 \eeq
while the corresponding velocity perturbation is governed by
 \beq
       \theta_K'  & = & (3w_K - 1){\cal H}\theta_K - w'_K{\theta_K\over 1 + w_K}
       \nn \\
       & & \mbox{}
       + {k^2c_K^2\delta_K\over 1 + w_K} + k^2\Psi. \label{fluct2}
 \eeq
Here the prime denotes the derivative with respect to conformal time, ${\cal H}
= a'/a$ and the equation of state parameter $w_K$ is given by (\ref{w_K}). Its
conformal time derivative is then
 \be
       w'_K = -{3{\cal H}q\nu^2 B \rho_c^{-q\nu}\over [1 + B\rho_c^{-q\nu}]^2}
 \ee
and goes to zero at late times. The remaining parameter is the Cardassian speed
of sound\footnote{With this definition $c^2$ coincides with $\delta p/\delta \rho$ only 
when perturbations are adiatibatic. In this paper we restrict to the adiabatic case.} 
$c_K^2 \equiv (\del p_K/\del\rho_K)_s = dp_K/d\rho_K$. From the pressure
(\ref{press}) we then find
 \be
      c_K^2 =  \frac{\nu B \rho_c^{-q\nu}[(\nu - 1) B \rho_c^{-q\nu}+q\nu -1]}{1+(2-\nu )B\rho_c^{q\nu}+(1-\nu)B^2 \rho_c^{-2q\nu}}.
 \ee
At late times when $\rho_c \ra 0$ this is seen to approach $c_K^2 \ra -\nu$ which is unacceptable. This can be avoided only by choosing the special
value $\nu = 1$. Then the expression for the sound speed simplifies to
\be
        c_K^2 = {q - 1\over 1 + \rho_c^q/B}.
\ee
In order for it to be positive, we must have $q > 1$ and also $q < 2$ so that $c_K^2 < 1$ at late times.

We find it also useful to do the calculations in the synchronous gauge, in which the line element is
given by \be
      d s^2 = a^2(\tau)[-d\tau^2 + (\delta_{ij} + h_{ij}) dx^i dx^j],
\ee
where we define the scalar modes in the Fourier space as
\be
       h_{ij}(\bx,\tau) & = & \int d^3k e^{i\bk \cdot \bx} \biggl[\hat{k}_i\hat{k}_jh(\bk,\tau) \nn \\
       &  & \mbox{} +  6(\hat{k}_i\hat{k}_j -
       \frac{1}{3}\delta_{ij})\eta(\bk,\tau)\biggr],
\ee
when $\bk=k\hat{k}$. In this gauge the fluctuation equations for a general fluid, which may have also anisotropic stress $\sigma$, are
\beq
       \delta' = -(1 + w)(\theta + \frac{h'}{2}) + 3\mathcal{H}(w -  c^2)\delta \label{fluct1s}
\eeq
for the density contrast and
\beq
       \theta' = (3w - 1)\mathcal{H}\theta - \frac{w'}{1+w}\theta + \frac{c^2}{1 + w}k^2\delta - k^2\sigma \label{fluct2s}
\eeq
for the velocity perturbation.
\begin{figure}[th]
\begin{center}
\includegraphics[width=0.45\textwidth]{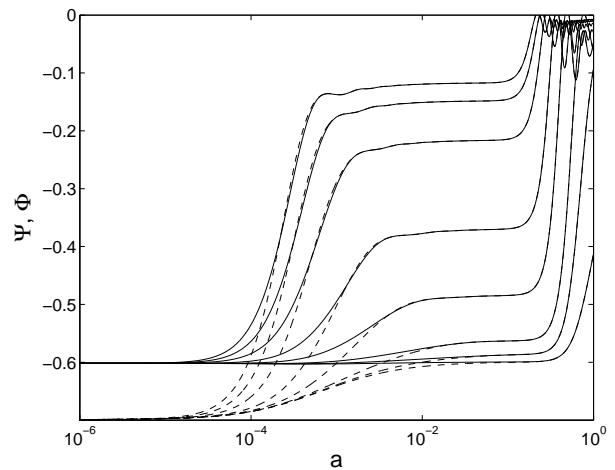}
\caption{\label{psi2}Evolution of $\Psi$ (solid line) and $\Phi$ (dotted line) in the
presence of a Cardassian fluid (case II). From bottom to top, $k = 0.00024,
0.0012, 0.0024, 0.0060, 0.012, 0.036,0.048$ Mpc$^{-1}$.}
\end{center}
\end{figure}

\begin{figure}[th]
\begin{center}
\includegraphics[width=0.45\textwidth]{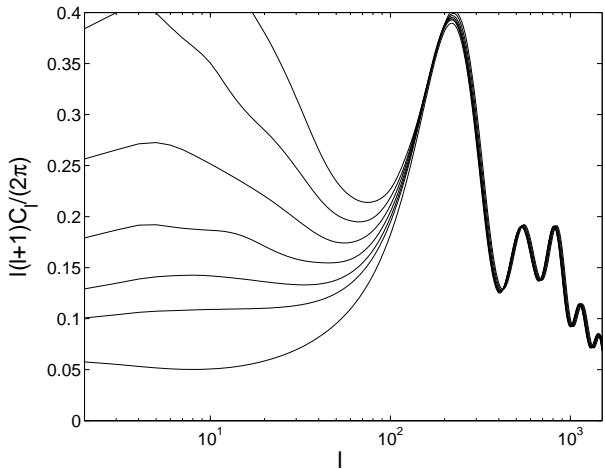}
\caption{\label{spectra2}CMB spectra in the case II. Here $\nu=1$ and from bottom to up, $q=1,1.05,1.1,1.2,1.4,1.8$ and $2.6$.}
\end{center}
\end{figure}

We have numerically solved the perturbation equations in both gauges now
including also the fluctuations in the Cardassian fluid for the representative
value $q = 1.5$. In this case the fluctuations in the dark matter are not
independent. The evolution of the Bardeen potentials $\Psi$ and $\Phi$ at
different scales is shown in Fig.\ref{psi2}. They are seen to have a very rapid increase
(decrease in absolute value) at late times for all scales comparable and
smaller than the present horizon. This results in an unusually strong late-time
ISW effect. In the thermal spectrum in Fig.\ref{spectra2} it is seen to give a rather big
deviation from the WMAP data for scales $\ell < 100$. We find the same results
in both gauges. The top curve is an example of a model where the sound speed of
the Cardassian fluid exceeds the speed of light. Models where $c_K^2$ takes
negative values are not computable at all since fluctuations then blow up, except for tiny
deviations from the $\Lambda$CDM values.

\begin{figure}[th]
\begin{center}
\includegraphics[width=0.45\textwidth]{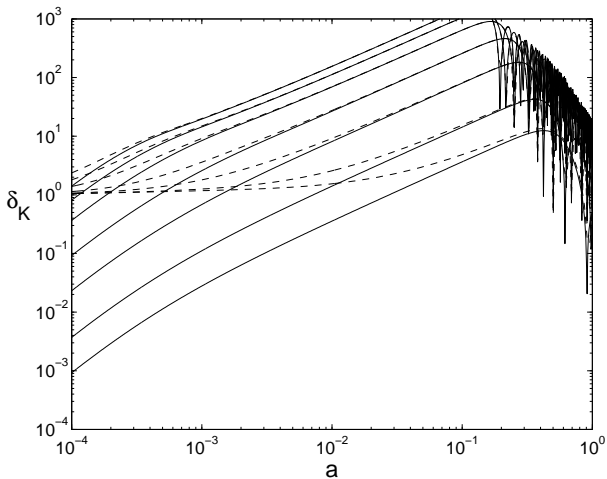}
\caption{\label{deltas}Evolution of $\delta_K$ in the case II. The model is the same as in Fig.4.
From bottom to top, $k = 0.0012, 0.0024, 0.0060, 0.012, 0.036,0.048$ Mpc$^{-1}$. Dashed lines represent the perturbations in Newtonian and the solid
lines in the synchronous gauge.}
\end{center}
\end{figure}

The dynamics is now different from the case I, where the growth of matter perturbations was slowed down only because of the accelerated expansion of
the background Universe. Now the overdensities, which have grown in the Cardassian fluid during its matter-like behaviour, are swiftly decaying as the
fluid is turning into an effective cosmological constant. Meanwhile, the decaying overdensities perform rapid oscillations due to presence of the
Cardassian pressure. This is shown in Fig.\ref{deltas}. Fluctuations in the cold dark matter and in the Cardassian fluid are related adiabatically
 \beq
      \delta_c = \left(\frac{d \log\rho_K}{d \log\rho_c}\right)^{-1} \delta_K
      = \frac{\delta_K}{1+w_K} \sim a^{3q} \delta_K, \quad \label{deltac}
 \eeq
where the last equality holds at late times (and when $\nu=1$). From this we see that the dark matter perturbations are driven along with the
Cardassian oscillations, but are not sharing the decay rate of the host fluid perturbations.

The Cardassian fluid which is restricted here to have the two parameters $\nu=1$ and $1<q<2$,
is related to the generalized Chaplygin gas (GCG)\cite{Bento:2002uh}. In the latter model the energy
density is
 \be
\rho_{GCG} = [B+\frac{A}{a^{3(1+\alpha)}}]^{\frac{1}{1+\alpha}}.
 \ee
After the change of variable, $q=1+\alpha$, we see that this reduces to the Cardassian model
in the special case that $\nu=1$ (A is fixed by requiring standard early cosmology). In the Cardassian model
we consider cold dark matter existing separately and that its perturbations obey Eq.(\ref{deltac}).
But the resulting CMB spectrum should be equivalent regardless of this intrinsic decomposition
of the fluid. Indeed, our results agree with\cite{Amendola:2003bz}, where the CMB spectrum was calculated
for the generalized Chaplygin gas. Note that there the
normalization of the spectra is different from ours, since there the Sachs-Wolfe plateau is
kept low, resulting in lower amplitude at small scales.

\begin{figure}[lt]
\begin{center}
\includegraphics[width=0.45\textwidth]{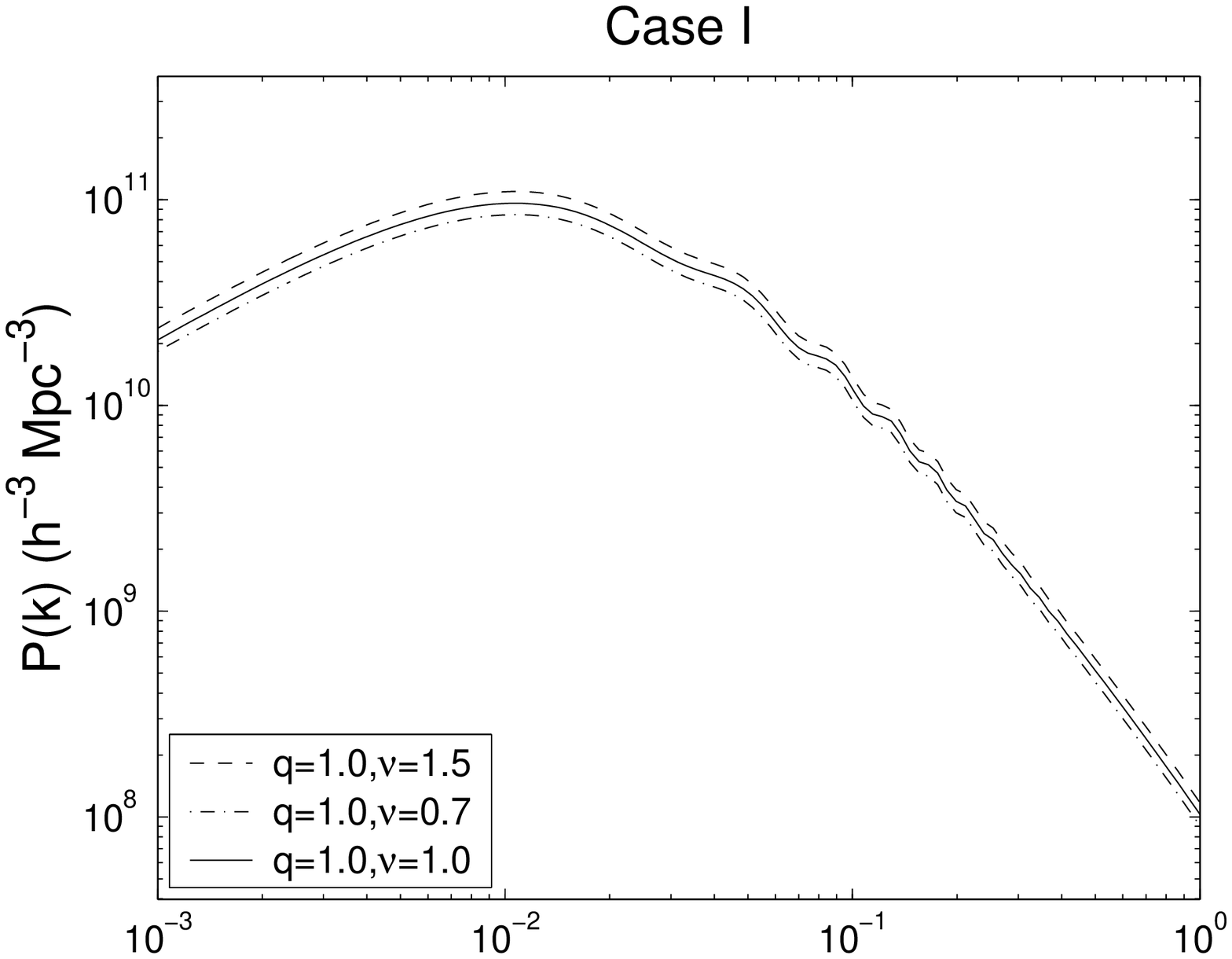}
\includegraphics[width=0.45\textwidth]{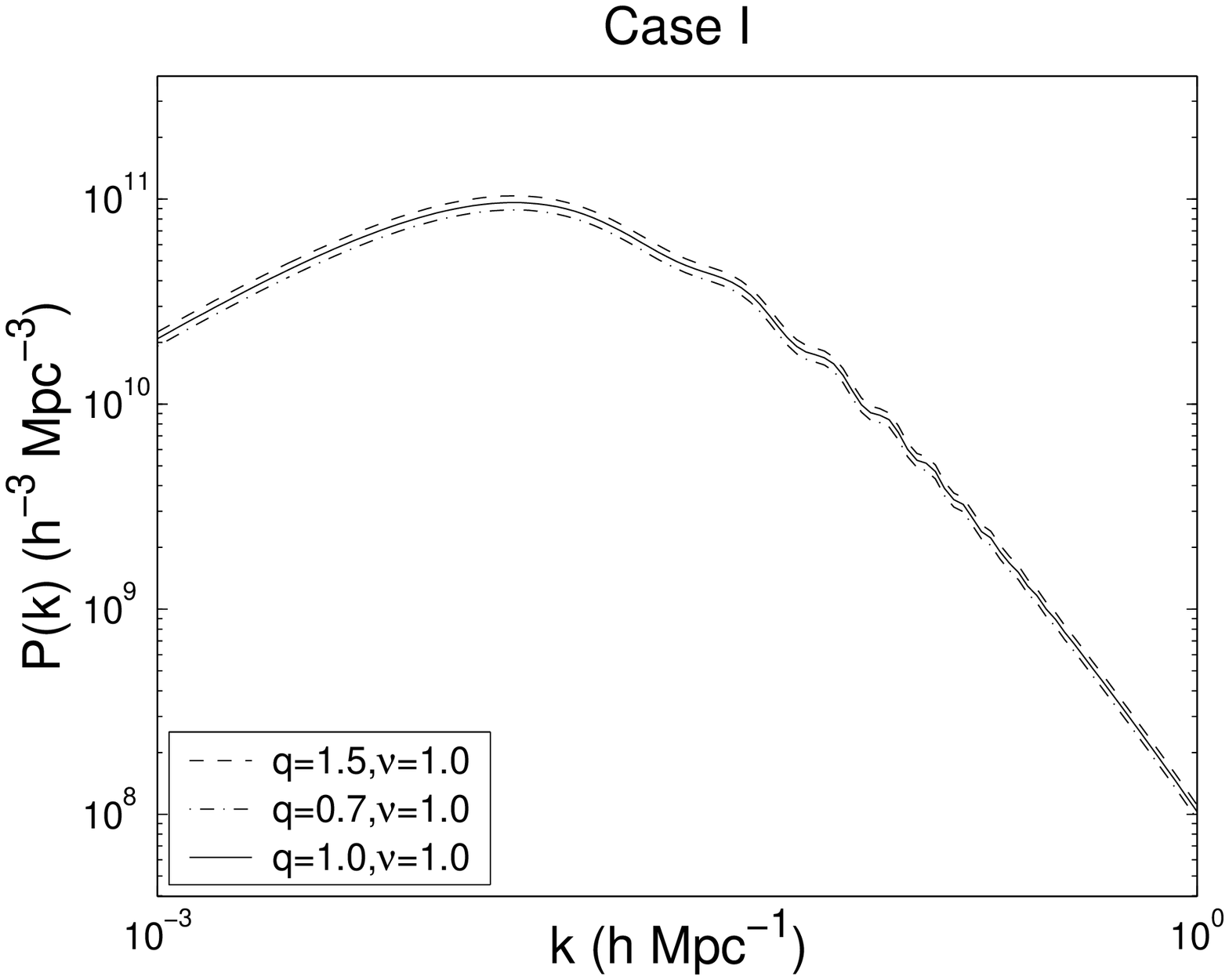}
\caption{\label{mat1}The total matter power spectra for different parameter values in case I.}
\end{center}
\end{figure}

\begin{figure}[rt]
\begin{center}
\includegraphics[width=0.45\textwidth]{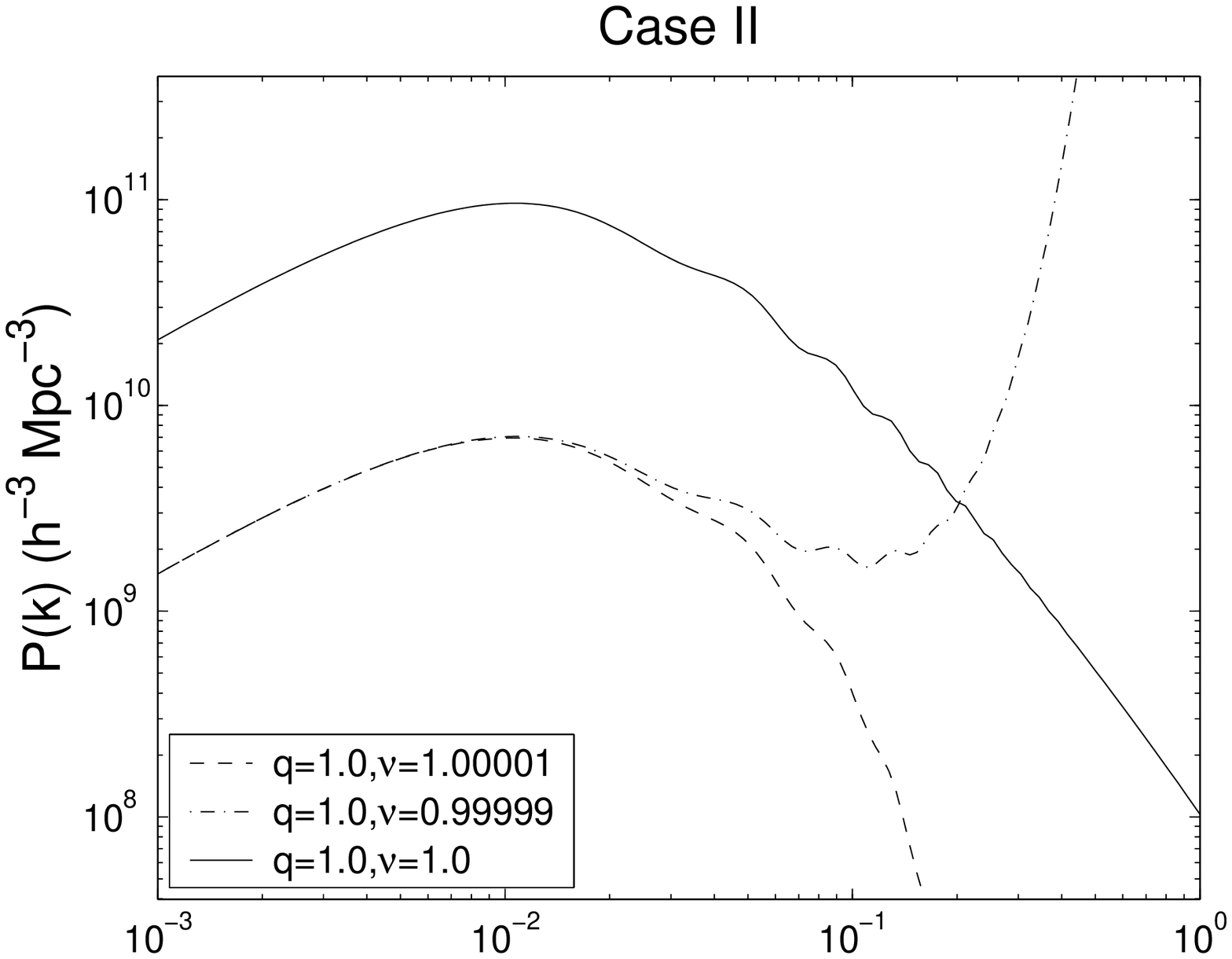}
\includegraphics[width=0.45\textwidth]{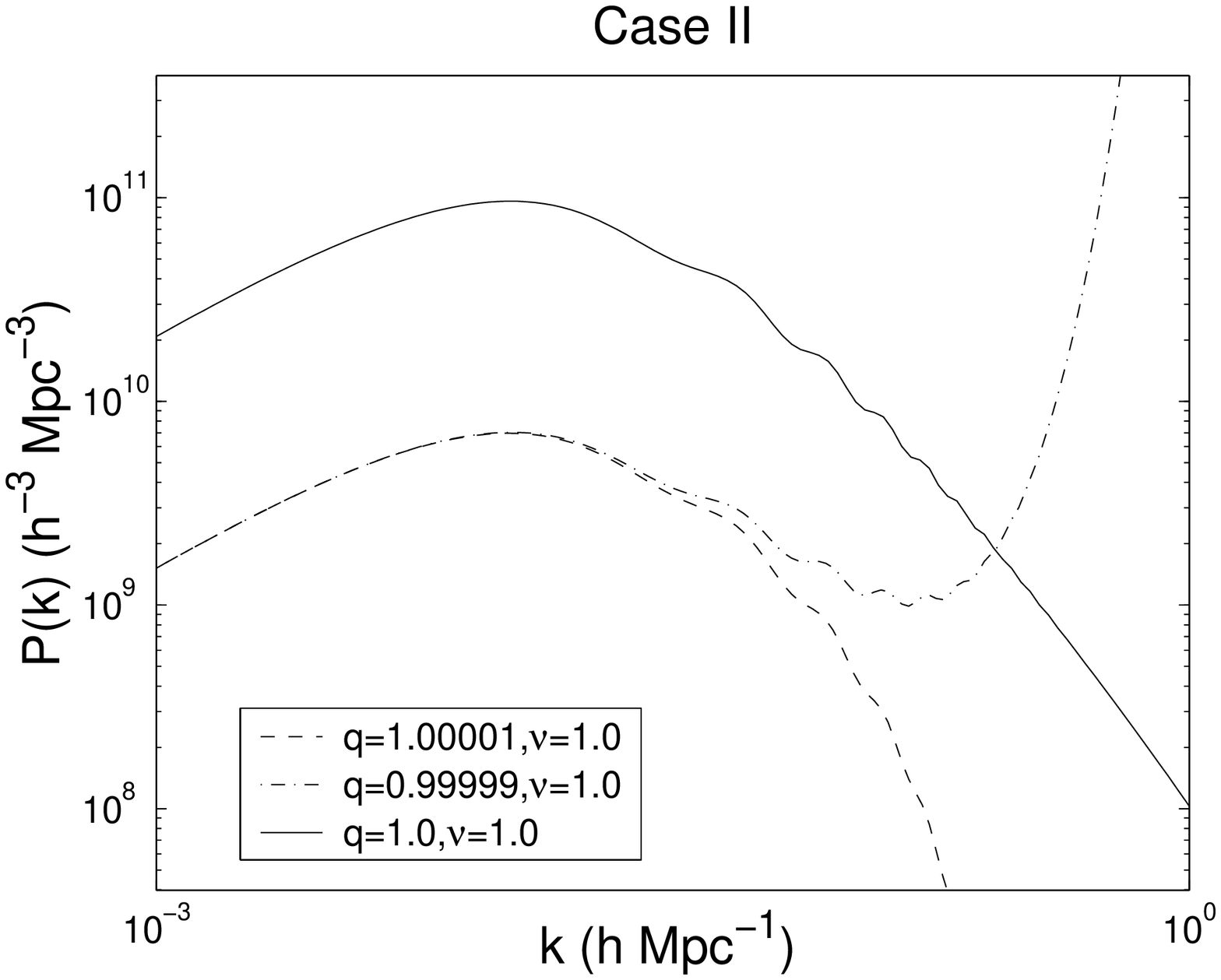}
\caption{\label{mat2}The total matter power spectra for different parameter values in case II.}
\end{center}
\end{figure}

For completeness, we consider also the matter power spectra in these models. We define
here the total matter power spectrum, including both the contribution from the Cardassian fluid and from the baryons, as

 \be
P(k) = \frac{k^3}{2 \pi^2}|\delta_k|^2 = \frac{k^3}{2 \pi^2} \left(\frac{\rho_K\delta_K+\rho_b\delta_b}{\rho_K+\rho_b}\right)^2.
 \ee
In the Fig.\ref{mat1} we see that modifying only the background evolution has little effect
on the matter power spectrum. However, as shown also by  Amarzguioui, Elgar\o y and Multam\"aki\cite{Amarzguioui:2004kc}, consistently
with  Sandvik, Tegmark and Zaldarriaga\cite{Sandvik:2002jz}, when one adopts the fluid interpretation of the Cardassian expansion,
even tiny changes in the parameters $q$,$\nu$ result in observable departure from the $\Lambda$CDM cosmology
in the matter power spectrum. This is seen in Fig.\ref{mat2}. When $c_S^2 < 0$, the spectrum blows
up. When $c_S^2 > 0$, the spectrum of Cardassian fluid would be oscillating as expected from Fig.\ref{deltas}, but these are
not seen in the total matter power spectrum since $\delta_K$ is suppressed and thus the baryonic fluctuations dominate the
structure at smaller scales. Also the amplitude of the total matter power spectrum is sensitive to the parameters $q$ and $\nu$.
There has been some interesting suggestions on how to render the growth of structure in the
GCG model more plausible\cite{Reis:2003mw,Bento:2004uh}. The one proposed in\cite{Reis:2003mw} imposes intrinsic entropy
perturbations in the fluid to cancel the effect of the finite sound speed. This could be also
done to the Cardassian fluid we are considering. In practice it would amount to simply setting
$c_K^2=0$ in Eqs.(\ref{fluct1}) and (\ref{fluct2}). However, since entropy perturbations are gauge-dependent,
it would perhaps be difficult to find physical justification for this. Another recent proposal\cite{Bento:2004uh} introduces an
unique decomposition of the Chaplygin gas into interacting vacuum energy and cold dark matter. In the case of
Cardassian fluid, the corresponding decomposition would require the redefinition of the cold dark matter density as
 \be
\tilde{\rho_c} & \equiv & \rho_K+p_K \nn \\
  & = &
\rho_c[1+B\rho_c^{-q\nu}]^{\frac{1}{q}-1}[1+(1-\nu)B\rho_c^{-q\nu}].
 \ee
Then we would find that the equation of state of the fluid $\rho_K-\tilde{\rho_c}$ would be a
constant, with $w=-1$. In\cite{Bento:2004uh} it was assumed that such a component would be unfluctuating and
that the redefined cold dark matter would have no pressure perturbations. However, these assumptions
are invalid if one is restricted to standard general relativity and adiabatic perturbations, since then
$\rho_K'\delta_c = \rho_c'\delta_K$, and now $\rho_K' \neq 0$ because of the interaction.

Both of these proposals, the imposement of entropy perturbations and the decomposition
of the fluid into dark matter and unperturbed vacuum energy, would prevent oscillations in the overdensities
and result in a smaller ISW effect for the Cardassian model than in Fig.\ref{spectra2}, but admittedly these
proposals seem rather ad hoc unless one could specify a physical reason responsible for neglecting
the undesiderable features in the evolution of inhomogenities in the fluid. Now we instead move on
to consider the possibility that the form of the Cardassian Friedmann equation arises not from energetics of
physical fluids but from modifications to the Einstein gravity.

\section*{Case III: Cardassian fluctuations induced by the CDM.}

To avoid the dark matter oscillations driven by the Cardassian pressure, one would have
to modify the scenario so that the dark matter would not see the Cardassian pressure. This
can be achieved, without relaxing the usual mass conservation or the adiabaticity of
the cold dark matter, if it would satisfy energy conservation separately. This would
remove any but the gravitational interaction between $\rho_c$ and $\rho_K - \rho_c$, and so there
should be no reason in ordinary physics why the parametric relation between the two should
be maintained. However, we have tried to motivate such a scenario by modified gravity as discussed
in the introduction.

Now both the matter and the Cardassian perturbations would obey the
corresponding Eqs.(\ref{fluct1s}) and (\ref{fluct2s}). In our case I they were
satisfied only for the dark matter, in our case II only for the Cardassian
fluid. It is not immediately clear that they can be satisfied for both
components simultaneously. This is easiest to see in a synchronous gauge, since
in such particular choice of gauge the velocity perturbations in matter, and
thus also in the Cardassian fluid can be set to zero. Thus we have only to
solve Eq.(\ref{fluct1s}) for dark matter,
 \beq
     \delta_c' = -\frac{h'}{2}, \label{fluct1c}
 \eeq
and invert Eq.(\ref{deltac}) to get the density perturbation of the Cardassian
fluid. To check that there indeed are no Cardassian velocity perturbations in
this gauge, one can rewrite Eq.(\ref{fluct1s}) as
 \be
     \theta_K = -\frac{h'}{2} - 3\mathcal{H}\frac{c_K^2 - w_K}{1+w_K}\delta_K - \frac{\delta_K'}{1+w_K}.
 \ee
Since by Eq.(\ref{deltac}) $\delta_K' = 3\mathcal{H}(w_K-c_K^2)\delta_K + (1+w_K)\delta_c'$, the RHS vanishes identically. Eq.(\ref{fluct2s}) tells
us that there now is anisotropic stress in the fluid, which is proportional to the density perturbation evaluated in the synchronous gauge:
\beq
     \sigma_K & = & \frac{c_K^2}{1+w_K}\delta_K = c_K^2\delta_c  \nn \\
            & = &    \frac{\nu B\rho_c^{-q\nu}[(\nu-1)B\rho_c^{-q\nu}+q\nu-1]}{[1+(1-\nu)B\rho_c^{-q\nu}]^2}\delta_K.\label{stress}
\eeq
Except for the last equality, this formulation applies generally for modified Friedmann equations with $\rho_K(\rho_c)$,
since it does not depend on the actual form of Eq.(\ref{cardass}).

Now the overdensities in pressureless dark matter give arise to perturbations
in the Cardassian density, in which the shear stress acts to eliminate the
effect of pressure gradients (and thus the sensitivity to $c_K^2$).
Thus no late-time oscillations occur in the overdensities of either component.
In this case one has the freedom to choose any positive value for $q$ and
$\nu$, since $\delta_K$ is now given by Eq.(\ref{deltac}) and thus $c_K^2$ can
take any value, even negative. The ISW effect is again strong, since now the
Cardassian fluctuations at late times are suppressed according to
Eq.(\ref{deltac}).

The stress term becomes important at late times, and it causes deviation
between the gravitational potentials in the Newtonian gauge. This is seen from
the constraint equation
 \be
     \Psi & = & \Phi - \frac{3}{2}\frac{a^2}{M^2k^2}(\rho+p)\sigma \\
     & = & \Phi - \frac{3}{2}\frac{a^2\nu B\rho_c^{-q\nu+1}}{M^2k^2}\frac{(\nu-1)B\rho_c^{-q\nu}+q\nu-1}{[1+B\rho_c^{-q\nu}]^{2-\frac{1}{q}}}
     \delta_c^\mathrm{(Syn)}. \nn
 \ee
\begin{figure}[!h]
\begin{center}
\includegraphics[width=0.45\textwidth]{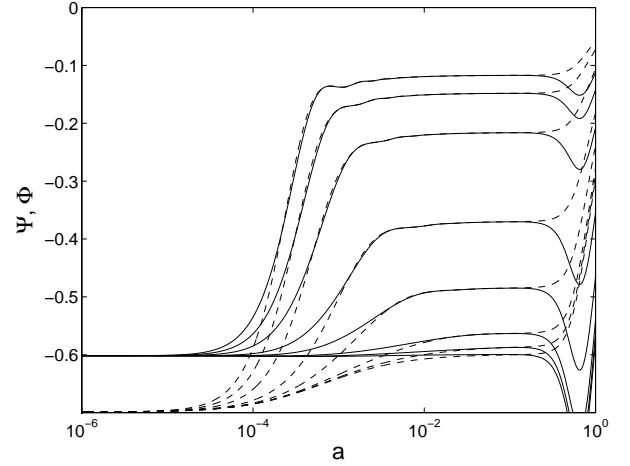}
\caption{\label{psi3}Evolution of $\Psi$ (solid lines) and $\Phi$ (dashed lines) in the
presence of the Cardassian fluid (case III). From bottom to top, $k = 0.00024,
0.0012, 0.0024, 0.0060, 0.012, 0.036,0.048$ Mpc$^{-1}$.}
\end{center}
\end{figure}
\begin{figure}[!h]
\begin{center}
\includegraphics[width=0.45\textwidth]{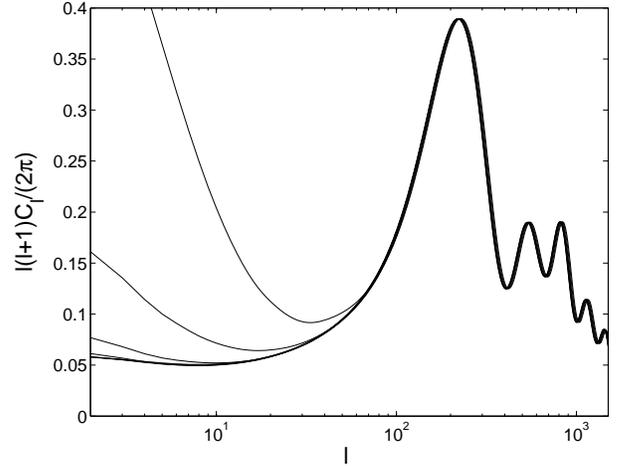}
\caption{\label{spectra3}CMB spectra in the case III. Here $\nu=1$ and from bottom to up, $q=1,1.05,1.1,1.2,1.4,1.8$ and $2.6$.}
\end{center}
\end{figure}
The $\sigma$ in the first line is the total gauge-invariant shear stress which at late times gets a contribution only from $\sigma_K$. In the
second line we have explicitly written down the relation of the shear stress to the density perturbation in the synchronous gauge, after
$\delta_K$ in terms of $\delta_c$ from Eq.(\ref{deltac}).

The stress appears during the Cardassian take-over period, and increases
temporarily the magnitude of the metric perturbation $\Psi$. When the
Cardassian density is beginning to dominate, the effect of stress is fading
away. The integrated Sachs-Wolfe  effect is stronger than in case I since now the effective matter
source of the gravitational potentials is $\rho_K\delta_K$ instead of $\rho_c\delta_c$.
The dip in the potential $\Psi$ may somewhat cancel the earlier contribution
to the integrated Sachs-Wolfe effect, but the net effect is, as in case II, that
the ISW again increases with increasing $q$. We show the evolution of gravitational
potentials in Fig.\ref{psi3} for parameters $(q,\nu)=(1.5,1)$. The CMB spectra is plotted in Fig.\ref{spectra3}
for the same parameter choices as we did in case II. Choosing parameter
values $\nu \neq 1$ seems to give similar results.

\begin{figure}[floatfix]
\begin{center}
\includegraphics[width=0.45\textwidth,height=0.4\textwidth]{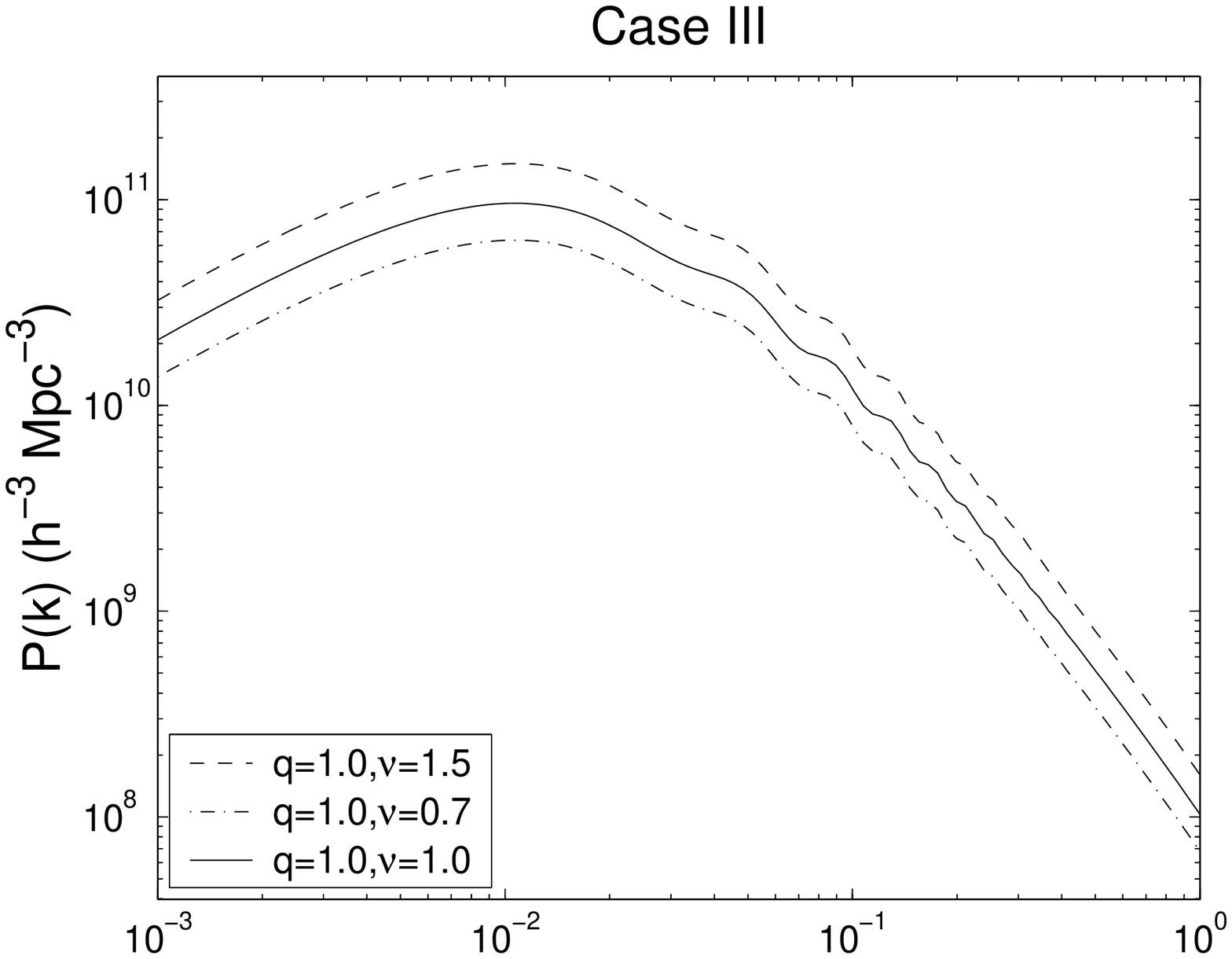}
\includegraphics[width=0.45\textwidth,height=0.4\textwidth]{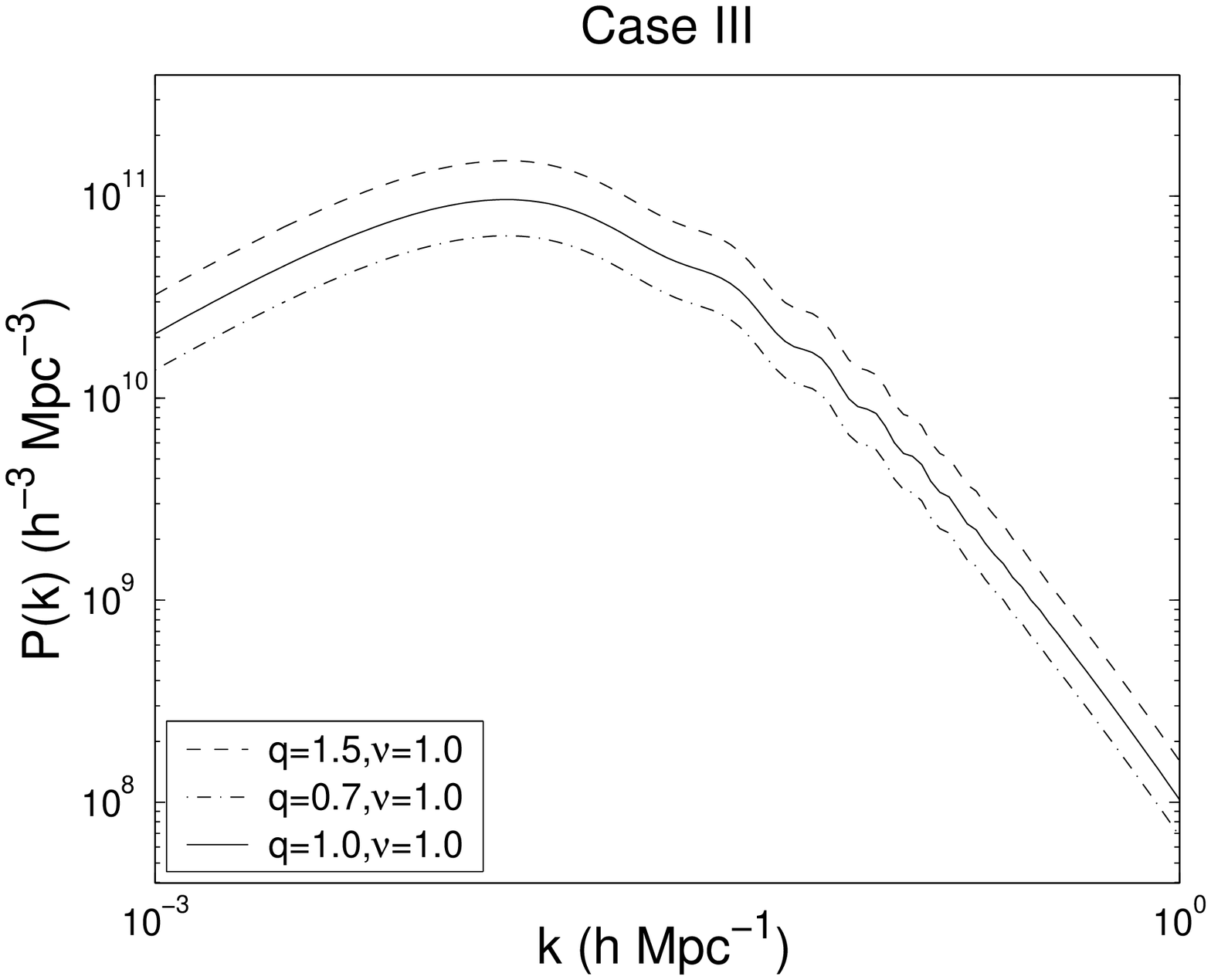}
\caption{\label{mat3}The total matter power spectra for different parameter values in the case III.}
\end{center}
\end{figure}
We plot the total matter power spectrum in Fig.\ref{mat3}. Evolution of the linear overdensities seems to be similar as in the Case I.
Thus the shape of the matter power spectrum is not as sensitive to the parameters $q$ and $\nu$ as the
CMB spectrum at large scales.

\section*{Conclusions}

We have attempted to describe modifications of the Friedmann equations in terms
of a cosmological fluid with a priori unknown properties. We applied different
assumptions on such a fluid to the representative case of the Cardassian model.
In case I we assume that the fluid has no fluctuations and just provides a
background for the evolution of the standard cosmological energy components.

A more physical assumption is considered in case $\rm{II}$ where we interpret the
Cardassian energy density as that of an interacting dark matter. It is shown that
then the requirement of a physically tolerable sound speed restricts the model
to the unmodified polytropic Cardassian one, which is equivalent to the
generalized Chaplygin gas model. These are known to predict damping of
the matter power spectrum\cite{Sandvik:2002jz} and CMB spectrum inconsistent with
observations\cite{Amendola:2003bz}. Following ideas introduced in the Chaplygin gas literature
\cite{Reis:2003mw},\cite{Bento:2004uh}, we comment on the possibility to improve the results
by modifying the fluid interpretation (of either Chaplygin gas scenario or the Cardassian
expansion).

In case III we interpret the additional energy associated with the dark matter
as arising from a modification to the general relativity. This is no longer
compatible with perfectness of the fluid represented by $T_{\mu\nu}^K$. When
Einstein equations are modified, we do not in general expect equality of the
Bardeen potentials in the absence of anisotropic stress. If we then write the
Einstein equations in the conventional form and take into account the
modifications in the form of a fluid, it is generally found to be imperfect. This
is also consistent with\cite{Lue:2003ky}, where modified Friedmann equations are
investigated under the assumption that the Birkhoff theorem is satisfied. Then
also $\Psi$ and $\Phi$ are found to differ. In fact our approach to modified gravity
in the case III coincides with the one in\cite{Lue:2003ky} at the scales they consider. Note that in the
formulation of our case III, there is no problem of
interpretation of the fluctuations in the present Universe. For the scales of interest,
the dark matter perturbations evaluated in different gauges do coincide at late times.
This is not true in the case II, as anticipated in\cite{Gondolo:2002fh}. Large scale structure
in the limit $k \rightarrow 0$ in these models with assumptions equivalent to our case 
III was studied also in\cite{Multamaki:2003vs}.

In both cases we find that the late integrated Sachs-Wolfe effect is typically
very large in the Cardassian models. For the fluid interpretation (case II) it is
clear, also from the matter power spectrum, that the model is not compatible with observations
unless one chooses parameters very close to the $\Lambda$CDM model. In the description
of modified gravity (case III) the matter power spectrum is not as drastically distorted when
$q,\nu \neq 1$, but it seems that the ISW effect could be used to rule out most of the parameter
space also in this case. However, a detailed likelihood analysis is beyond the scope of the present
paper.

Hannestad and Mersini-Houghton\cite{Hannestad:2004ts} have recently presented results of a more
general investigation on the effects of new physics on the CMB spectrum. They also find that modifications of the
late-time gravitational interactions can lead to a similar boost of the
low-$\ell$ part of the spectrum as seen here in the Cardassian model for our
cases II and III. It remains to be seen how generic this effect is and what
constraints it puts on the equation of state for the dark energy.

\begin{acknowledgments} We thank F. Finelli and N. Bilic for pointing out the relation to the GCG.
This work has been supported by NorFA and grant 159637/V30 from the Research Council of Norway. T.K. is grateful
to Waldemar von Frenckell Foundation and Emil Aaltonen Foundation for financial support and to University of
Oslo for hospitality.
\end{acknowledgments}

\bibliography{refs}

\end{document}